\documentclass[12pt]{iopart}

\usepackage{graphicx}
\usepackage{epstopdf}
\usepackage{epsfig}
\usepackage{graphics}
\usepackage{amssymb,amsfonts,graphics,color}
\usepackage{subfigure}
\usepackage{textcomp}
\usepackage{array}
\usepackage{multirow}
\usepackage{iopams}  
\usepackage{dcolumn}
\usepackage{bm}
\usepackage{setstack}
\usepackage{xr}
\usepackage{hyperref}
\newcommand{\bra}[1]{\ensuremath{\left\langle#1\right|}}
\newcommand{\ket}[1]{\ensuremath{\left|#1\right\rangle}}

\begin{document}

\title[Multiplexed on-demand storage of polarization qubits in a crystal]{Multiplexed on-demand storage of polarization qubits in a crystal}

\author{Cyril Laplane, Pierre Jobez, Jean Etesse, Nuala Timoney, Nicolas Gisin and Mikael Afzelius}

\address{Group of Applied Physics, University of Geneva, CH-1211 Geneva 4, Switzerland}
\ead{cyril.laplane@unige.ch}
\vspace{10pt}
\begin{indented}
\item[]\today
\end{indented}

\begin{abstract}
A long-lived and multimode quantum memory is a key component needed for the development of quantum communication. Here we present temporally multiplexed storage of 5 photonic polarization qubits encoded onto weak coherent states in a rare-earth-ion doped crystal. Using spin refocusing techniques we can preserve the qubits for more than half a millisecond. The temporal multiplexing allows us to increase the effective rate of the experiment by a factor of 5, which emphasizes the importance of multimode storage for quantum communication. The fidelity upon retrieval is higher than the maximum classical fidelity achievable with qubits encoded onto single photons and we show that the memory fidelity is mainly limited by the memory signal-to-noise ratio. These results show the viability and versatility of long-lived, multimode quantum memories based on rare-earth-ion doped crystals.
\end{abstract}

%
%
%
%
%

\section{Introduction}
\label{sec:Intro}
Distribution of entanglement over long distances is one of the main challenges of quantum information \cite{Gisin2007a}. For quantum communication or distributed quantum computing, one of the main pillars is the implementation of quantum memories. Indeed, such a device could serve as a light-matter interface between computing nodes and channels of communication \cite{Kimble2008}, moreover it is an essential building block of quantum repeaters \cite{Briegel1998,Duan2001,Sangouard2011}. These memories need to be efficient, multimode, long-lived and have high fidelities \cite{Lvovsky2009,Simon2010,Bussieres2013}.\\
\indent
For information technology, a multimode capacity of communication channels presents   the obvious advantage of increasing the density of information processed (computed or transmitted). In quantum information science, the information is often encoded, processed and carried by single quanta, for instance single photons. For quantum communication single photons present the drawback of being intrinsically sensitive to losses, which implies that any operation will be rather inefficient. One can overcome this problem by using multimode encoding. Indeed the multimode capacity of the memory can lead to multiplexing of operations thus effectively increasing the rate of, for instance quantum repeaters \cite{Collins2007,Simon2007} (i.e. the effective success probability of the operation). Depending on the system or the type of operation wanted, multiplexing can either be temporal \cite{Nunn2008,Usmani2010,Bonarota2011a}, spatial \cite{Higginbottom2012,Zhou2015,Nicolas2014} or spectral \cite{Sinclair2014}.\\
\indent
Ensemble-based quantum memories couple strongly to light due to their intrinsic collective enhancement effect.
Rare-earth ions doped crystals are promising candidates for quantum memories \cite{Tittel2010b,RiedmattenAfzeliusChapter2015} thanks to their large  bandwidth \cite{Thiel2014} and long coherence times \cite{Longdell2005,Heinze2013,Zhong2015} at cryogenic temperature. They present atomic-like properties without the need of heavy experimental apparatus for manipulation. Numerous achievements in the past years have shown the potential of these lanthanides activated solids for quantum memories. With the goal of achieving a functioning and scalable quantum memory with these crystals, several milestones have already been reached: high efficiency of 69$\%$ \cite{Hedges2010}, cavity enhanced storage for a preprogrammed delay \cite{Sabooni2013} and for spin-wave (i.e. on-demand) storage \cite{Jobez2014}, high multimode capacity \cite{Usmani2010,Bonarota2011a}, long storage time \cite{Longdell2005,Heinze2013}. Recently the longest coherence time ever (6 hours) was measured \cite{Zhong2015} in the same type of crystal we are using here. 
On-demand spin-wave storage of classical time-bin pulses has been achieved in \cite{Gundogan2013,Timoney2013} and recently time-bin qubits were successfully stored on-demand at the single photon level \cite{Gundogan2015}. Recently we also demonstrated on-demand spin-wave storage of a pulse at the single photon level for about 1 ms, using spin-echo techniques to extend the storage time \cite{Jobez2015}. The storage of polarization qubits has however been demonstrated for preprogrammed delays only \cite{Clausen2012,Gundogan2012,Zhou2012}.\\
\indent
Here we show the versatility of spin-wave memory at the single-photon-level by storing 5 polarization qubits multiplexed in time for more than half a millisecond (50 times longer than the previously cited work on qubit storage in crystals). Our device is based on an atomic frequency comb (AFC) spin-wave memory in an Eu$^{3+}$:Y$_2$SiO$_5$ crystal. We use the inherent temporal multiplexing to AFC memories to boost the rate of operation of our device by increasing the number of modes stored. We measure the fidelity of the retrieved qubits for several mean numbers of input photons and show that it is always above the classical bound of 2/3 for true single photons. Furthermore we can model the fidelity of our memory based solely on its phase coherence, efficiency and unconditional noise floor. This allows us to conclude that its fidelity is mainly limited by its signal-to-noise ratio (SNR). \\

\section{Memory scheme and experimental setup}
\label{sec:Setup}

\subsection{The AFC memory scheme}

The storage protocol we employ is the atomic frequency comb \cite{Afzelius2009a} quantum memory. We here give a brief description of the essential features of the protocol, particularly the time sequence of the storage process. For a more complete description we refer to \cite{RiedmattenAfzeliusChapter2015,Afzelius2009a}.

Taking advantage of the optical inhomogeneously broadened absorption, we tailor an atomic frequency comb in the absorption profile through a fine selective optical pumping. The peaks of the comb are separated by $\Delta$ which is much smaller than the input bandwidth. When a photon is absorbed, the single excitation is delocalized over all the atoms. This large collective state will quickly lose its collective coherence due to the frequency detuning of each atom with respect to the central frequency of the input photon. Since all the participating atoms are located in periodic positions in frequency space (i.e. $m\Delta$ with $m$ an integer) they will rephase automatically at a time $1/\Delta$ and thereby cause an echo-type reemission. We can intuitively see the reemission process as a temporal diffraction (the inverse Fourier transform of the spectral grating i.e. the comb) of the input pulse. As such the protocol is a delay line with a preprogrammed reemission time. In order to implement on-demand reemission and which can also give a longer storage time, we can transfer the optical coherence created by an absorbed photon into a spin coherence using a strong $\pi$-pulse called control pulse (see figure \ref*{fig:setupfinal}(b) and (c)). This will freeze the phase evolution due to the tailored inhomogeneous frequency profile. When we want to retrieve the stored input, a second control pulse is applied, which will restore the optical coherence and the AFC phase evolution will continue up to its completion, with the emission of an echo. This scheme is called an AFC spin-wave memory. Note that as shown in figure \ref*{fig:setupfinal}(c) each temporal modes spend the same time in the memory (both in the excited state and in the spin state), such that the storage efficiency is the same for each mode.  \\
\indent
As for the inhomogeneously broadened optical transition, the spin transition is broadened (27 kHz in our case). Using spin-echo techniques we can compensate the spin dephasing and thus achieve an order of magnitude longer storage time. The RF-sequence used in this work is called an XY-4 sequence \cite{Maudsley1986,Souza2011,AliAhmed2013a}. This sequence has the particularity of being robust to errors in population inversion, independent of the phase of the initial spin state. This is relevant in our experiment since the generation of the spin state through optical storage has a fluctuating phase. Indeed in a storage experiment at the quantum level, it is crucial to achieve very high quality of refocusing therefore avoiding unwanted population in the target spin states, which would cause fluorescence (i.e. noise) in the output mode (see \cite{Jobez2015}). Furthermore each of the four $\pi$-pulses is chirped in amplitude and frequency (adiabatic pulses \cite{Silver1985}) in order to achieve efficient population inversion over the entire 27 kHz bandwidth of the spin ensemble.\\
\indent
 The number of temporal modes that can be stored is given by the number of teeth in the comb \cite{Afzelius2009a}. The number of teeth can be increased either by increasing the bandwidth (for a fixed $1/\Delta$ storage time) or by increasing the $1/\Delta$ storage time (for a fixed bandwidth). One has to take extra care of the bandwidth of the input modes, which cannot exceed the bandwidth of the AFC. For the full spin-wave storage scheme, room has to be made for the transfer pulses and of course these control pulses must have the required bandwidth \cite{Jobez2015b}.\\

\subsection{Storing polarization qubits in a birefringent and anisotropically absorbing crystal}

\begin{figure*}
	\centering
    \includegraphics[width=1\textwidth]{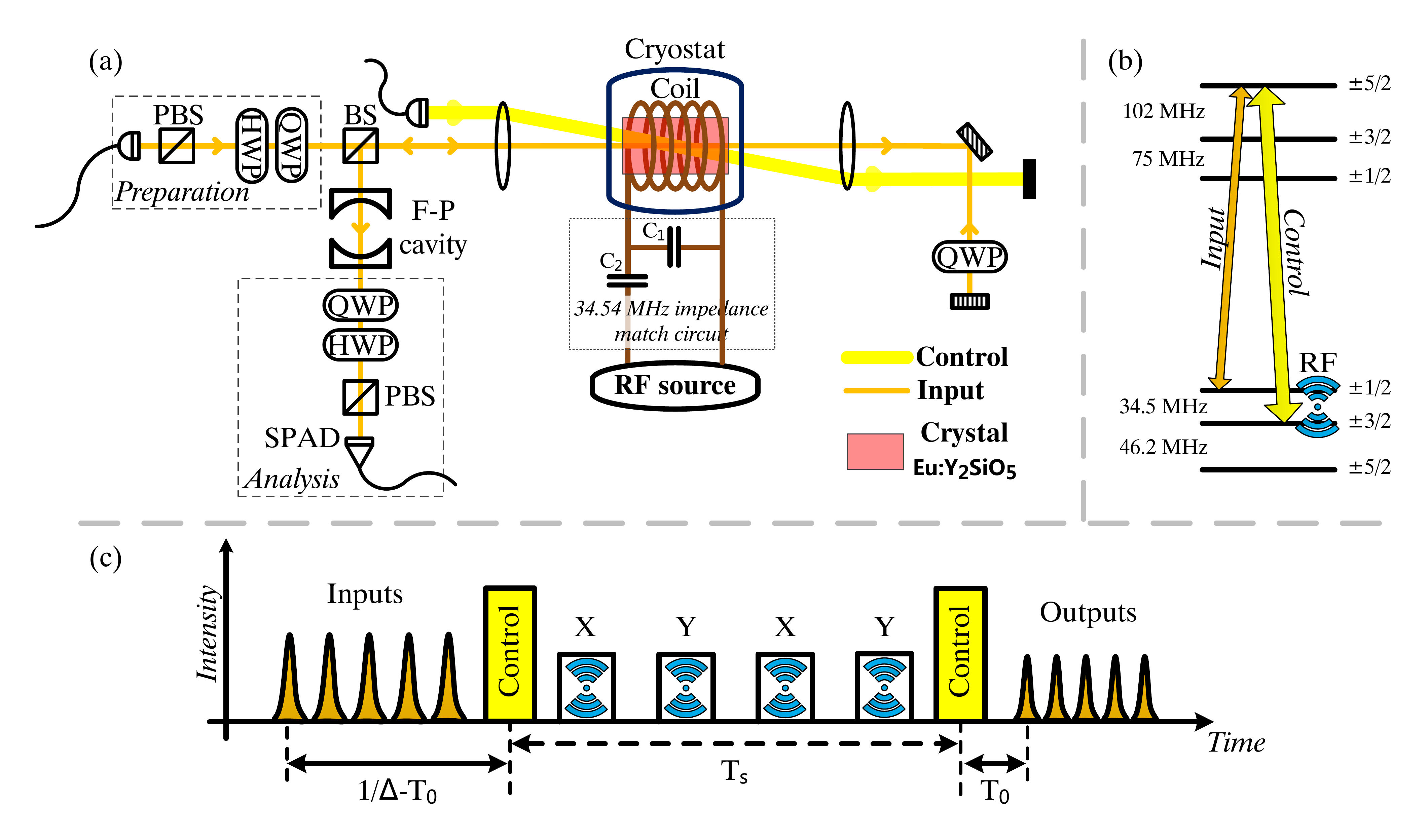}
        \caption{AFC spin-wave storage of polarization qubits. (a) Experimental setup. PBS: polarization beam splitter. BS: beam splitter. HWP: half-wave plate. QWP: quarter-wave plate. SPAD: single-photon avalanche photodiode. F-P: Fabry-Perot. (b) The hyperfine states of the ground and excited states and the two transitions of the chosen $\Lambda$-system in Eu$^{3+}$:Y$_2$SiO$_5$. (c) Timing of the storage sequence. The total storage time is $1/\Delta+T_S$, where $T_S$ is the spacing between the control pulses. The RF spin-echo sequence is inserted in between the optical control pulses. The RF pulses are applied using a coil placed around the crystal with a resonant circuit at 34.5 MHz (see (a)).}
 	\label{fig:setupfinal}
\end{figure*}

The Eu$^{3+}$:Y$_2$SiO$_5$ crystal is a birefringent crystal which exhibits anisotropic absorption. The crystal is cut along the plane (D1, D2) and the light is propagating along the b axis, where (D1, D2, b) are the principal dielectric axes. There is a coincidence between the principal axes of refraction and absorption so the highest absorption is obtained for light polarized along D1, the lowest for light with polarization parallel to D2 \cite{Koenz2003}. One can thus decompose an arbitrary polarization state into the orthonormal basis formed by D1 and D2.  Inspired by \cite{Clausen2012,Zhou2012} we are using a unique crystal in double pass configuration and a quarter-wave plate in-between the two passes (see figure \ref{fig:setupfinal}(a)). This ensures that both components of the polarization will experience the same absorption and global refraction. The polarization is preserved up to the swap operation induced by the quarter-wave plate in double pass configuration. We quantify the remaining anisotropy by measuring the optical depth and efficiency for different polarization inputs. We found that the optical depth varies of $\pm 15\%$ from the mean value which results in a variation of $\pm 9\%$ in the efficiency. 

\subsection{Experimental details}

\noindent \textbf{The Crystal and optical setup.}
We use an isotopically pure 1cm long europium doped yttrium orthosilicate crystal, Eu$^{3+}$:Y$_2$SiO$_5$, with an europium concentration of 1000 ppm. We work on the $^7$F$_0\rightarrow ^5$D$_0$ transition of site 1 at 580.04 nm whose inhomogeneous broadening is approximately 1.6 GHz and the overall absorption coefficient is $\alpha$=2.6 cm$^{-1}$ \cite{Ferrier2015}. The electronic ground and excited states have hyperfine structures \cite{Yano1991,Yano1992a} which are shown in figure \ref{fig:setupfinal}(b). The inhomogeneous spin linewidth of the 34.54 MHz spin transition is 27 kHz \cite{Jobez2015}. The crystal is cooled down below 4 K in a low-vibration cryostat (Montana Instruments Cryostation).\\
To produce the necessary yellow light, we use a diode laser at 1160 nm, frequency stabilized on a high finesse optical cavity under vacuum. The laser is frequency doubled and amplified to produce more than 1.8 W at 580 nm. 
The light is then sent through a series of acousto-optic modulators (AOMs) in order to prepare three modes: the input mode, the control mode and the stabilization mode. The optical setup is shown in figure \ref{fig:setupfinal}(a), where both the control and input modes are represented. The input mode is an attenuated coherent state, whose average photon number can be varied and whose initial polarization is adjusted with the help of a half and quarter waveplate (\textit{Preparation} in figure \ref{fig:setupfinal}(a)). The input mode is then sent through the crystal in the previously mentioned double-pass configuration. After this it is filtered by a Fabry-Perot cavity (finesse 400 and bandwidth 2.5 MHz) whose resonance is stabilized on the input frequency using the stabilization mode (not shown). The output mode is finally analyzed with the help of a quarter and half waveplate followed by a polarization beam splitter (\textit{Analysis} in figure \ref{fig:setupfinal}(a)). Note here that we are using only one port of the PBS for detection. The filtering is necessary due to the high power of the control mode ($\sim$ 10$^{15}$ photons) which creates fluorescence and off-resonant coherent emission (see \cite{Timoney2013}). To further enhance the filtering, the input and control modes are not applied in the same spatial mode. We detect the output using a single photon avalanche diode (SPAD) gated with another AOM (not shown in figure \ref{fig:setupfinal}(a)) before coupling into a fiber with a 9$\mu$m core. The total transmission from the input of the cryostat to the detector is 7$\%$. We measure an SPAD efficiency of 57$\pm$3$\%$ at 580 nm, with a dark count rate of 15 Hz. By measuring the counts at the detector and compensating for the rate of experiment, total transmission and detector efficiency we can infer the number of photons at the input of the memory. In the following we will always refer to this value.
\\
\noindent \textbf{AFC preparation.}
The atomic level scheme is shown in figure \ref{fig:setupfinal}(b). Through optical pumping, we first prepare a selected sub-ensemble of ions into the ground state $\vert\pm\frac{1}{2}\rangle_g$ \cite{Jobez2014}.
Second we create a 2 MHz wide AFC structure on the $\vert\pm\frac{1}{2}\rangle_g$ $\rightarrow$ $\vert\pm\frac{5}{2}\rangle_e$ transition using a 0.5 s long sequence of precise spectral holeburning \cite{Jobez2015b}. During the whole step we optically pump the $\vert\pm\frac{3}{2}\rangle_g$ $\rightarrow$ $\vert\pm\frac{5}{2}\rangle_e$ transition in order to keep the $\vert\pm\frac{3}{2}\rangle_g$ level empty, readying it for spin-wave storage. 
The entire preparation sequence is described in more detail in \cite{Jobez2014,Jobez2015b}.
\\
\noindent \textbf{Storage sequence}
The full storage sequence is shown in figure \ref{fig:setupfinal}(c): five input modes of 1.25 $\mu$s each are sent in the crystal and stored with the help of two control pulses of 5 $\mu$s duration each. In between the optical control pulses, four population-inverting (i.e. $\pi$ rotation) RF pulses resonant with the spin transition are applied with a periodic spacing of $T_S$/4. Each pulse has a duration of 120 $\mu$s and has a total frequency chirp range of 45 kHz around 34.54 MHz. They are generated with an arbitrary function generator and sent to a 100 W amplifier. The amplified signal is then sent through a RF circulator, an impedance matched circuit made of two tunable capacitors (see in figure \ref{fig:setupfinal}(a)) and finally a 7-turn coil wrapped around the crystal. The impedance-matching circuit is tuned to resonate at 34.54 MHz, allowing us to achieve a Rabi frequency of 47 kHz.\\
We measured a transfer efficiency of 70$\pm$5\% per optical control pulse. The absorption probability of the input mode is measured to be 70$\pm$3\% (see table \ref{TAB:fidelity_modetr} in appendix C) for the comb setting $1/\Delta=15\ \mu$s, such that the total conversion efficiency from the optical mode to the spin-wave is about 50\%. The total storage time is $1/\Delta+T_S$ = 515 $\mu$s.\\
In the experiment we repeat this storage sequence $N_{rep}=18$ times, with an experiment rate of 0.8 Hz, which leads to an average rate of storage sequence of 14 Hz. By proceeding so, we considerably increase the effective rate of the experiment. This repetition increases the average population in $\vert\pm\frac{3}{2}\rangle_g$, which in turn would increase the unconditional noise probability $p_n$ \cite{Timoney2013} with increasing number of repetitions $N_{rep}$. To reduce this accumulation of population we apply a repump pulse on the $\vert\pm\frac{3}{2}\rangle_g$-$\vert\pm\frac{5}{2}\rangle_e$ transition after each storage cycle.\\
In the present experiment, we store and retrieve five temporally multiplexed photonic qubits encoded with the same state for more than half a millisecond. The different stored states are $\vert$H$\rangle$, $\vert$V$\rangle$, $\vert$D$\rangle$=$\frac{1}{\sqrt{2}}$($\vert$H$\rangle$+$\vert$V$\rangle$) and $\vert$R$\rangle$=$\frac{1}{\sqrt{2}}$($\vert$H$\rangle$+i$\vert$V$\rangle$) with different average number of input photons ($\mu=0.8,$ $1.4,$ $3.6,$ $8.2$).
\\

\section{Theoretical model and experimental results}

We first characterize the fidelity of the memory with bright coherent pulses ($\mu>>1$). This allows us to work in a regime where the noise created by the storage process is negligible in comparison with the signal. We can then evaluate the phase coherence of the memory by calculating its classical fidelity, in order to determine its effect on the total conditional fidelity.
This classical fidelity can be inferred from the visibility of a classical echo signal:

\begin{equation}
V_c=\frac{S_{max}-S_{min}}{S_{max}+S_{min}},
\label{Vis} 
\end{equation}

\noindent
where $S_{min}$ and $S_{max}$ are the classical signals measured respectively after an orthogonal and parallel analyzer with respect to the input. The fidelity is then given by 

\begin{equation} 
F_c=\frac{1+V_c}{2}=\frac{S_{max}}{S_{max}+S_{min}}.
\label{Fid_c}  
\end{equation}

\noindent
Using pulses prepared in various states and equation \ref{Fid_c} we calculate a mean classical fidelity of 99.1 $\pm$ 0.4 $\%$. This shows the good coherence preservation of the memory.

\begin{figure}[h]
    \centering
    \includegraphics[width=1\textwidth]{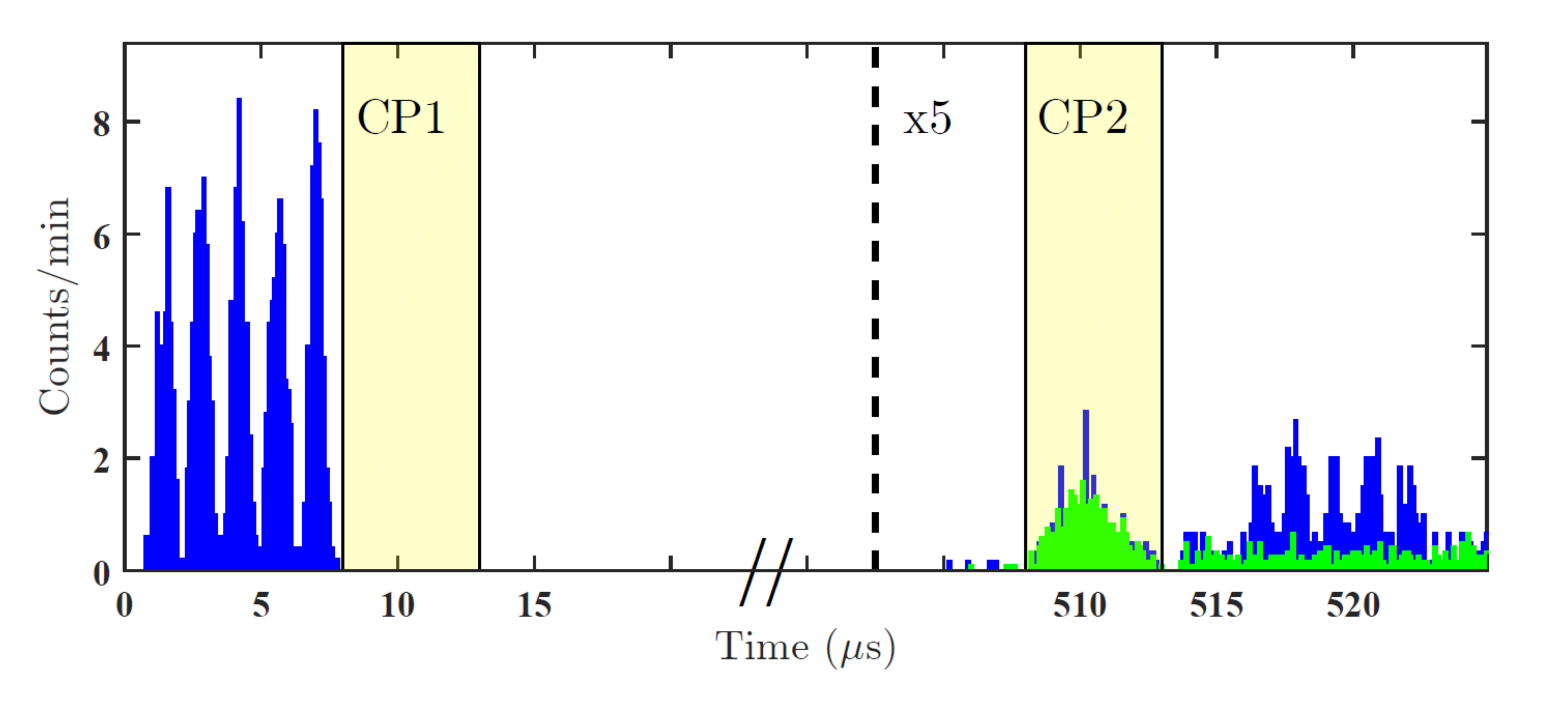}
    \caption{Photon counting histogram measured for a $\vert$D$\rangle$ encoded input with $\mu$=1.4. The 5 input and output pulses are shown in blue, starting at 0 and 515 $\mu$s, respectively. The output is measured in the $\vert$D$\rangle$ basis. The noise produced by the memory (the unconditional noise floor) is shown in green, which was measured with no input pulse but still implementing the entire memory protocol. The CP1 and CP2 regions show where the 5 $\mu$s long control pulses are applied. Note the leakage of the control pulse through the AOM gate in region CP2.}
    \label{summary}
\end{figure}

Now that we have shown that the memory preserves the states encoded on bright pulses, we lowered the number of photons in the input to investigate the behaviour of the memory in the single photon regime. Indeed in this regime the noise produced by the memory reduces the visibility of the retrieved state. Figure \ref{summary} shows the photon counting histogram obtained by storing $\vert$D$\rangle$=$\frac{1}{\sqrt{2}}$($\vert$H$\rangle$+$\vert$V$\rangle$) with $\mu$=1.4 and measured in the $\vert$D$\rangle$ basis. Similar temporal histograms are obtained when measuring in the $\vert$A$\rangle$, $\vert$H$\rangle$ (or $\vert$V$\rangle$) and $\vert$R$\rangle$ (or $\vert$L$\rangle$) bases. We then calculate the fidelity via tomographic state-reconstruction using a maximum likelihood method. The experiment is repeated with different mean number of photons $\mu$ in the input.
In order to assess the quality of the multimode capacity of the AFC, we can check the individual conditional fidelities of the five stored photonic qubits. The fidelities vary from 83.3 $\pm$ 3.8$\%$ to 86.6 $\pm$ 2.9 $\%$ for $\mu$=1.4, showing that the storage fidelity over the different temporal modes is preserved (see table \ref{TAB:fidelity_mode} in appendic A).

We now investigate how the fidelity varies as a function of the mean photon number in the input. We also use the multimode capacity in order to increase the rate of the experiment, as it would be the case in a quantum repeater \cite{Bussieres2013}. The counts of the 5 modes are summed to achieve statistics that would otherwise take 5 times (the number of modes stored) longer to accumulate. The results are summarized in table \ref{TAB:fidelity_vs_n}.

\begin{table}[h]
	\begin{center}
	\begin{tabular}{l|l|l|l|l}
		  	$\mu$ & $\eta$ (\%)  & $p_n$ ($10^{-3}$) & $\mu_1$ & Fidelity (\%) \\ 
		  	\hline
0.8 $\pm$ 0.1 & 4.3 $\pm$ 0.4  & 11.0 $\pm$ 1.0 & 0.25 $\pm$ 0.04 & 79.5 $\pm$ 0.2  \\
1.4 $\pm$ 0.1 & 3.6 $\pm$ 0.3  & 10.1 $\pm$ 1.2 & 0.28 $\pm$ 0.04 & 85.5 $\pm$ 0.1  \\
3.6 $\pm$ 0.3 & 3.8 $\pm$ 0.2  & 10.9 $\pm$ 1.4 & 0.29 $\pm$ 0.04 & 93.6 $\pm$ 0.1  \\
8.2 $\pm$ 0.6 & 3.7 $\pm$ 0.2  & 12.1 $\pm$ 1.4 & 0.33 $\pm$ 0.05 & 95.7 $\pm$ 0.04  \\

	\end{tabular}
	\end{center}
		\caption{Conditional fidelities measured for various mean input photon numbers $\mu$ prepared in $\vert$D$\rangle$. The fidelity was found via tomographic state-reconstruction using a maximum likelihood method, and the errors are estimated via Monte-Carlo simulation. Shown are also the storage efficiency $\eta$, the unconditional noise floor $p_n$ and the $\mu_1$ parameter (see text).}
		\label{TAB:fidelity_vs_n}
\end{table}

\begin{figure}[h]
	\centering
	\includegraphics[width=1\textwidth]{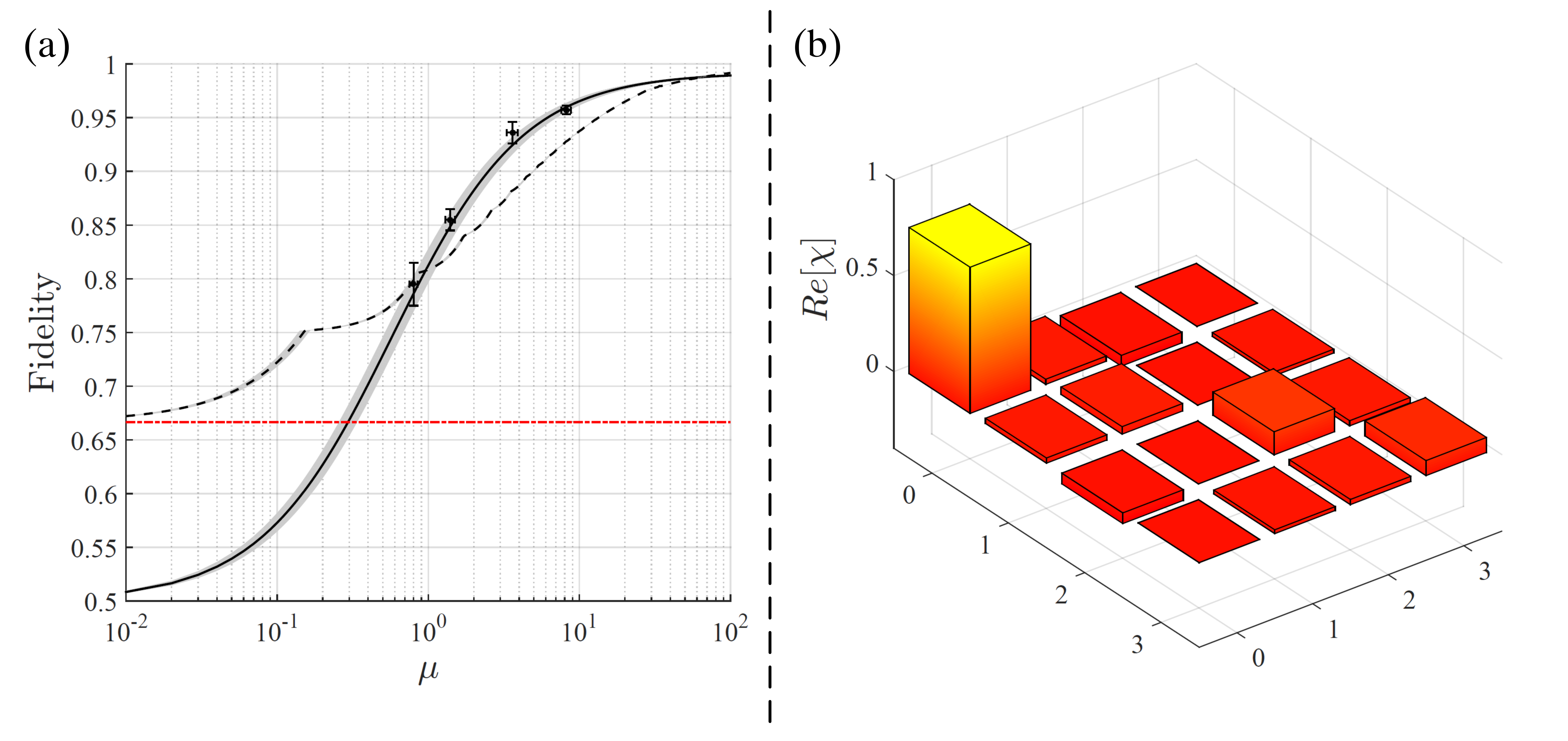}
	\caption{(a) Conditional fidelity as a function of the mean number of photons in the input. The dashed red line represents the maximum achievable fidelity with a classical memory and single photons. The black curve (grey area) is a plot of equation (\ref{Fid_q2}) for $\mu_1$=0.29 ($\pm$ 0.04) and $F_c$=99.1 $\%$. The black dashed curve is the maximum achievable fidelity with a classical memory and taking advantage of the non-unit efficiency and the non-definite (Poissonian) number of photons intrinsic of coherent states (see appendix D for more details). (b) Real part of the process matrix as obtained from the quantum process tomography.}
	\label{Fidel_vs_N_wcrit}
\end{figure}

The measured fidelity is reduced as the mean photon number is lowered. To understand this behaviour we derive an equivalent of equation (\ref{Fid_c}) for the fidelity $F_q$ in the case where the stored state is at the single photon level. An additional parameter has to be considered in this regime: the unconditional noise floor $p_n$ \cite{Timoney2013}. It corresponds to the probability of recovering a photon in the output time window after storage without any input, and allows for the definition of a new parameter $\mu_1=p_n/\eta$ where $\eta$ is the memory efficiency. The $\mu_1$ parameter is the mean photon number in the input that gives a SNR of 1 in the ouput. Using this figure of merit, we can estimate the fidelity of the memory assuming that the noise of the memory is state independent \cite{Gundogan2015,Jobez2015}. As shown in table \ref{TAB:fidelity_1,4} in appendix B this asumption is valid in our case. The fidelity at the single photon level can be calculated using equation (\ref{Fid_c})

\begin{equation}
F_q=\frac{S_{max}^{q}}{S_{max}^q+S_{min}^q}, 
\label{Fid_q} 
\end{equation}

\noindent
where $S_{max}^q$=$\mu\eta F_c$+$p_n$ and $S_{min}^q$=$\mu\eta (1-F_c)$+$p_n$, given that the input contains $\mu$ photons. Then

\begin{equation}
F_q=\frac{\eta\mu F_c+p_n}{\eta\mu +2p_n}=\frac{F_c+\frac{\mu_1}{\mu}}{1+2\frac{\mu_1}{\mu}}.
\label{Fid_q2} 
\end{equation}
\noindent
This formula is valid for a weak coherent state with a mean photon number $\mu$ as well as for a single photon where $\mu$ (with $0<\mu<1$) would be the probability to find the photon before the memory. In figure \ref{Fidel_vs_N_wcrit} we compare the fidelities predicted by equation (\ref{Fid_q2}) and the measured ones. The agreement is excellent for all mean photon numbers, and underscores that the fidelity is limited by noise produced by the memory. We note that a higher efficiency would lead to much higher fidelities for mean photon numbers below 1. \\

\indent
The fidelities in table \ref{TAB:fidelity_vs_n} are always above the classical limit of 2/3 valid for single-photon qubit storage \cite{Massar1995}, showing that our device is capable of faithfully storing single photons with high SNR. However as it has been shown in \cite{Gundogan2012} and \cite{Specht2011}, one can also characterize the quantum nature of a device using weak coherent states. To that end we have to consider the maximum storage fidelity achievable for coherent states with a classical storage device. For instance a classical memory could measure the input, try to deduce its state based on the so gained information, store the result and finally prepare and resend a new state (a measure-and-prepare strategy). The fidelity it could reach for a single-photon qubit is limited to 2/3 \cite{Massar1995}, but in the case of a coherent state, the mimicking device could extract more information for states containing more photons. Furthermore such a classical storage device could take advantage from the non-unit efficiency of the memory by sending a signal only when the number of photons in the input is above a certain threshold, which would enable it to reach even higher fidelities \cite{Gundogan2015,Specht2011}(see figure \ref{crit_newgen} in appendix D).

Here we additionally consider the fact that our device does not fully absorb the input state, thereby producing a transmitted signal in addition to the output signal. This means that a classical storage device should in principle reproduce the transmitted signal as well. As explained in further detail in appendix D the fidelities attainable by the classical counterpart are lower than the measure-and-prepare strategy on the output alone. This is because the transmission efficiency is higher than the output efficiency. This new criterion is plotted in figure \ref{Fidel_vs_N_wcrit}(a) (dashed line) and with respect to it, our memory can be called quantum for the inputs with $\mu$= 1.4, 3.6, 8.2 but we cannot conclude with certainty using the state containing $\mu = 0.8$ photons on average in the current state of the experiment. Nevertheless we would like to emphasize that the memory was developed with the goal of storing single photons, which we proved to be feasible provided that the probability of finding the photon before the memory is higher than $\mu_1=0.29$.   

Finally, we fixed the mean number of photons to be $\mu$=1.4 and we probed our memory with the states $\vert$H$\rangle$, $\vert$V$\rangle$, $\vert$D$\rangle$ and $\vert$R$\rangle$ in order to check if the memory can store any of those states in the single photon regime. Two methods were used to characterize the memory. The first method simply consisted in calculating the average fidelity, which was found to be 84.1 $\pm$ 0.1 $\%$, with low deviation from one state to the other (see table \ref{TAB:fidelity_1,4} in appendix B). The second method consisted in constructing the process matrix $\chi$ associated with our device \cite{Nielsen2000}. It is defined as follows: if we input an arbitrary state $\rho_{in}$ the memory will output the state $\rho_{out}$ defined by

\begin{equation}
 \rho_{out}=\sum_{k,l=0}^{3}\chi_{kl}\sigma_k\rho_{in}\sigma^\dagger_l,
 \label{process_mat_eq}
\end{equation}

\noindent
where $\sigma_{k,l}$ are the Pauli matrices. \\
The memory is perfect if and only if $\chi_{00}$=1 and all the others contributions are zero: in this case $\rho_{out}$=$\rho_{in}$. The real part of the reconstructed $\chi$ matrix is shown in figure \ref{Fidel_vs_N_wcrit}(b) indicating the good robustness of our memory ($\chi_{00}$=0.762) which is limited by the SNR. However, phase and bit-flip errors are present in the process ($\chi_{22}$ and $\chi_{33}$ errors), which possibly could be explained by a residual anisotropic absorption as a function of polarization. Indeed, our method for compensating the anisotropy is not perfect. A weak residual anisotropy will render the efficiency of the memory weakly polarization dependent hence the SNR will slightly vary from one state to the other. We think that with further work this polarization absorption dependence can be reduced. Another approach would be to increase the efficiency of the memory hence its fidelity, as already mentioned above.

\section{Conclusions and Outlook}
\label{sec:Conclusions_Outlook}
We have shown that the AFC spin-wave memory protocol can store multiple qubits at the single photon level. We have shown faithful storage of five temporally multiplexed polarization qubits for more than half a millisecond in a rare-earth-ion doped crystal. Our work shows that the spin-wave manipulation used for storage time extension does not degrade the fidelity of the stored states, paving the way for very long storage times of quantum states beyond the spin coherence time through dynamical decoupling. The excellent agreement of the measured fidelities with those predicted by our simple model allows us to foresee the possibility of very high fidelities (a crucial prerequisite for quantum repeaters) if we increase enough the efficiency using for instance an impedance matched cavity around our crystal \cite{Jobez2014}.

\section*{ACKNOWLEDGEMENTS}
The authors thank Alban Ferrier and Philippe Goldner for the crystal fabrication. We would like to thank Alexey Tiranov, Anthony Martin and Jonathan Lavoie for useful discussions, as well as Raphael Houlmann and Claudio Barreiro for technical support. This work was financially supported by the Swiss National Centres of Competence in Research (NCCR) project Quantum Science Technology (QSIT), by the European projects SIQS (FET Proactive Integrated Project) and CIPRIS (People Programme (Marie Curie Actions) of the European Union Seventh Framework Programme FP7/2007-2013/ under REA Grant No. 287252).

\newpage
\section*{APPENDIX A: Results of the individual temporal modes}
\label{sec:suppl_info_indiv_mod}
Here we take advantage of the multimode capacity of our memory by storing 5 temporal modes all encoded with the same qubit ($\vert$D$\rangle$). The following table shows that the conditional fidelity is the same for all five temporal modes:

\begin{table}[h]
	\begin{center}
	\begin{tabular}{c|c|c|c|c|c}
		  	mode & $\mu$ & $\eta$ (\%)  & $p_n$ ($10^{-3}$) & $\mu_1$ & Fidelity (\%) \\ 
		  	\hline
1 & 1.2 $\pm$ 0.1 & 3.5 $\pm$ 0.6  & 8.8 $\pm$ 1.3 & 0.25 $\pm$ 0.08 & 84.9 $\pm$ 3.6  \\
2 & 1.5 $\pm$ 0.1 & 4.3 $\pm$ 0.6  & 12.0 $\pm$ 1.5 & 0.28 $\pm$ 0.07 & 86.6 $\pm$ 2.9  \\
3 & 1.5 $\pm$ 0.1 & 3.2 $\pm$ 0.5  & 9.0 $\pm$ 1.4 & 0.28 $\pm$ 0.08 & 86.4 $\pm$ 3.5  \\
4 & 1.5 $\pm$ 0.1 & 3.5 $\pm$ 0.6  & 10.5 $\pm$ 1.4 & 0.30 $\pm$ 0.08 & 85.7 $\pm$ 3.2  \\
5 & 1.5 $\pm$ 0.1 & 2.6 $\pm$ 0.5  & 9.4 $\pm$ 1.2 & 0.36 $\pm$ 0.10 & 83.3 $\pm$ 3.8  \\

	\end{tabular}
	\end{center}
	\caption{Conditional fidelities of the 5 individual temporal modes, each encoded as $\vert$D$\rangle$. Shown are also (for earch mode) the average number of photons $\mu$, in storage efficiency $\eta$, the unconditional noise floor $p_n$ and the $\mu_1$ parameter (see main text).}
	\label{TAB:fidelity_mode}
\end{table}

\section*{APPENDIX B: Results of the different encoded states}
\label{sec:suppl_info_diff_state}
We give here the measured conditional fidelities for differently encoded input states. These fidelities were used to calculate the process fidelity of the memory. We see here that the noise $p_n$ is not dependent on the input state.
\begin{table}[h]
	\begin{center}
	\begin{tabular}{l|l|l|l|l}
		  	$\vert$$\Psi_{in}$$\rangle$ & $\eta$ (\%)  & $p_n$ ($10^{-3}$) & $\mu_1$ & Fidelity (\%) \\ 
		  	\hline
$\vert$H$\rangle$ & 3.3 $\pm$ 0.3  & 9.3 $\pm$ 1.3 & 0.28 $\pm$ 0.05 & 84.1 $\pm$ 0.2  \\
$\vert$V$\rangle$ & 3.7 $\pm$ 0.3  & 12.3 $\pm$ 1.5 & 0.33 $\pm$ 0.05 & 84.0 $\pm$ 0.1  \\
$\vert$D$\rangle$ & 3.6 $\pm$ 0.3  & 10.1 $\pm$ 1.2 & 0.28 $\pm$ 0.04 & 85.5 $\pm$ 0.1  \\
$\vert$R$\rangle$ & 3.1 $\pm$ 0.2  & 11.3 $\pm$ 1.6 & 0.36 $\pm$ 0.06 & 82.6 $\pm$ 0.1  \\

	\end{tabular}
	\end{center}
		\caption{Conditional fidelities measured for a mean input photon number $\mu$=1.4$\pm$0.1 prepared into various input states $\vert$$\Psi_{in}$$\rangle$. Shown are also the storage efficiency $\eta$, the unconditional noise floor $p_n$ and the $\mu_1$ parameter (see main text).}
		\label{TAB:fidelity_1,4}
\end{table}

\newpage
\section*{APPENDIX C: Results for the transmitted states}
\label{sec:suppl_info_trans_state}
We here show the measured transmission coefficients for the five temporal input modes and their measured conditional fidelities. These numbers were used to calculate the criterion plotted in figure \ref{Fidel_vs_N_wcrit}(a) (black dashed curve).
\begin{table}[h]
	\begin{center}
	\begin{tabular}{c|c|c}
		  	mode & Transmission (\%)  &  Fidelity (\%) \\ 
		  	\hline
1 & 33.8 & 97.2 $\pm$ 0.4 \\
2 & 28.0 & 96.8 $\pm$ 0.5  \\
3 & 30.4 & 97.4 $\pm$ 0.4  \\
4 & 30.1 & 97.6 $\pm$ 0.4  \\
5 & 25.5 & 97.0 $\pm$ 0.5  \\

	\end{tabular}
	\end{center}
	\caption{Conditional fidelities of the individual transmitted temporal modes measured for an input encoded in $\vert$R$\rangle$ and $\mu=1.4$.}
		\label{TAB:fidelity_modetr}
\end{table}

\section*{APPENDIX D: A quantum behaviour of the memory}
\label{sec:Etesse_criterion}
In this section, we derive the criterion that was used to infer a quantum behaviour of our memory, given that we probe it with coherent states only.\\
We first summarize an already existing criterion \cite{Gundogan2012}, which only takes into account the non-unit efficiency of the memory in terms of the output. We then modify the criterion by also considering that a less than 100 \% efficient memory cannot entirely absorb the input state. The main idea relies on the fact that the memory transmits part of the input state with a very good fidelity, so that a classical device that would like to reproduce the behaviour of the memory should also reproduce this transmitted state.
\subsection{``Measure-and-prepare strategy'' on the output only}
To infer a quantum behaviour of our memory, we propose to show that a particular protocol based on a measure-and-prepare strategy cannot mimic the experimental results. In other words, if we suppose that instead of the memory we place an eavesdropper, traditionally named Eve, that has only classical storage means at his disposal, we would like to know how well she could simulate the experimental results. It is known that with this strategy and an input qubit encoded into exactly $n$ photons, the best achievable fidelity is \cite{Massar1995}:
\begin{equation}
F=\frac{n+1}{n+2}.
\label{MP}
\end{equation}
In our experiment we do not use Fock states with a well defined number of photons $n$ but we use a coherent state with a mean photon number $\mu$, so that equation (\ref{MP}) has to be modified. In theory, Eve could take advantage of the coherence of the state to optimize her measurements on the state and maximize the fidelity of the states she re-sends. It is though implicitly assumed that the experimentalist that prepares the state can easily blur the phase of the coherent state so that the stored state is only a statistical mixture of Fock states, encoding the polarization onto states of the form:
\begin{equation}
\hat{\rho}_{in}=\sum_{n=0}^{\infty}P(\mu,n)\ket{n}\bra{n},\qquad\mbox{with}\qquad P(\mu,n)=e^{-\mu}\frac{\mu^n}{n!}.
\end{equation}
Then, the output conditional fidelity is simply the statistical mixture of the fidelities for $n\geqslant1$ photons:
\begin{equation}
F_M(\eta)=\sum_{n=1}^{+\infty}\frac{n+1}{n+2}\ \frac{P(\mu,n)}{1-P(\mu,0)}=\frac{1}{1-e^{-\mu}}\Big[\frac{1-e^{-\mu}-\mu+\mu^2}{\mu^2}-\frac{e^{-\mu}}{2}\Big].
\end{equation}
\noindent Note that since we use photon counting techniques we do not measure the vacuum state $\ket{0}$, hence the sum in the equation above starts at $n=1$. This is then by definition a conditional fidelity, i.e. conditioned on the detection of at least one photon.

\subsection{Efficiency consideration}
In \cite{Gundogan2012,Specht2011}, the authors have pushed the considerations one step further by taking into account the output inefficiency of the memory, so that Eve does not need to output a state each time. Eve can use this to only send an output state only if the number of photons is high enough in the input ($n\geqslant n_{min}$). The conditional fidelity is then higher, and can be written as:
\begin{equation}
 F_M(\mu,\eta_M)=\frac{\displaystyle\frac{n_{\rm min}+1}{n_{\rm min}+2}\ \gamma+\displaystyle\sum_{n\geqslant n_{\rm min}}\frac{n+1}{n+2}P(\mu,n)}{\gamma+\displaystyle\sum_{n\geqslant n_{\rm min}}P(\mu,n)},
\label{fidelbase}
\end{equation}
where $0<\gamma<P(\mu,n_{min})$ is a parameter that can be adjusted to mimic the memory efficiency $\eta_M$. 

\subsection{``Measure-and-prepare strategy'' on the output and the transmitted states}
\subsubsection{Idea of the strategy\\}

\begin{figure}[!h]
\begin{center}
\includegraphics[width=15cm]{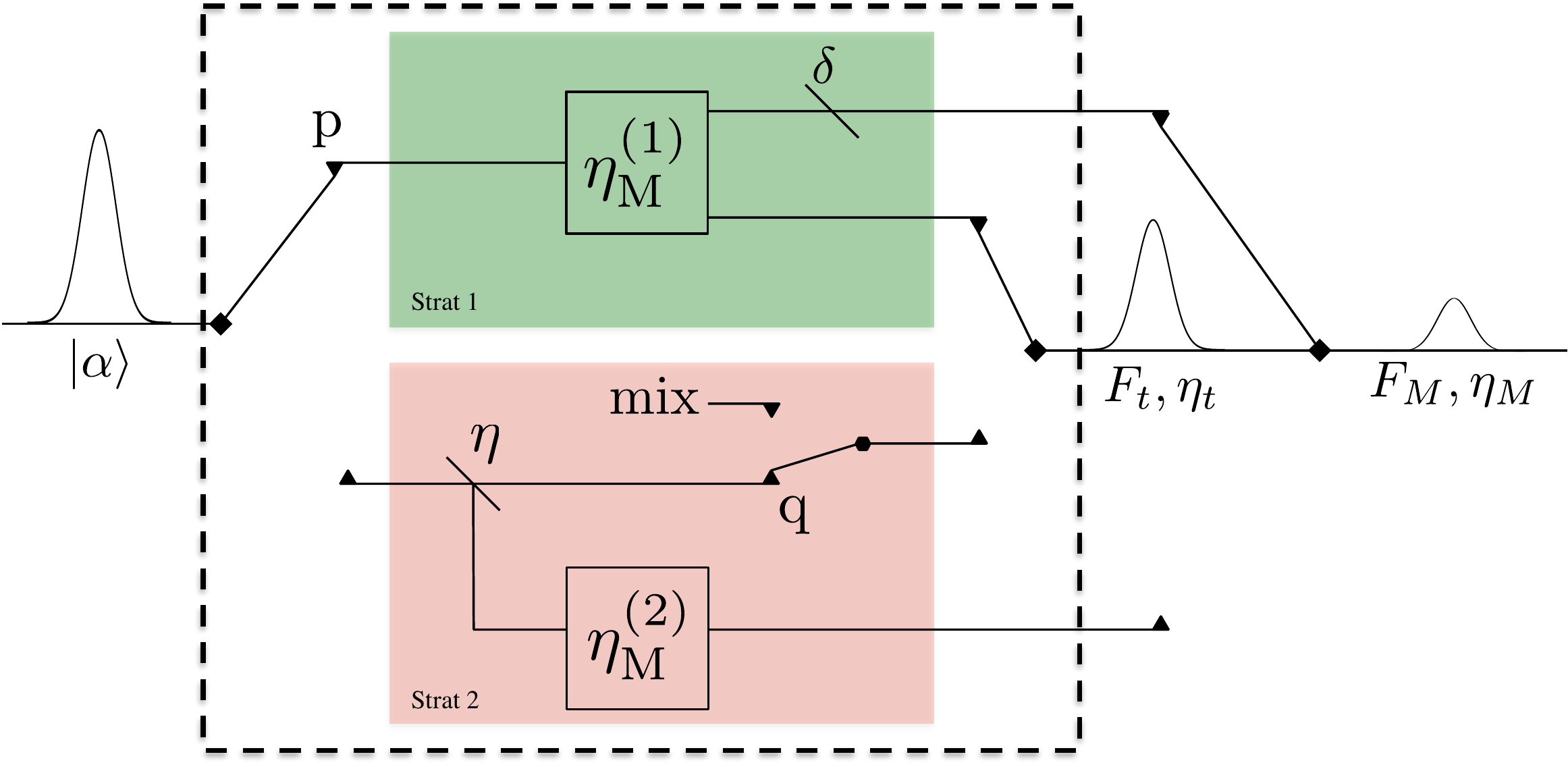}
\caption{A classical scheme for imitating the quantum memory. The two measure-and-prepare strategies are shown in the two boxes $\eta_{\rm M}^{(1)}$ and $\eta_{\rm M}^{(2)}$.}
\label{fideltrans}
\end{center}
\end{figure}

In order to duplicate the complete behaviour of the memory, a classical scheme should reproduce its imperfections. One of these is the fact that the coherent state to be stored is not fully absorbed by the material, and part of it leaks into a transmitted state. We propose to implement a possible scheme based on the previously introduced classical measure-and-prepare strategies, shown on figure \ref{fideltrans}.

In our analysis, we will suppose that whatever the strategy of Eve, she has to send a transmitted pulse which has exactly the same fidelity $F_t$ and efficiency $\eta_t$ as the ones of the experiment, and the stored output pulse should be emitted with the same overall memory efficiency $\eta_M$ as the one of the experiment. The parameter that she will try to maximize in the scheme is the fidelity of the output state $F_M$. The conclusion of a possible quantum behaviour of the memory will finally be inferred by comparing the best achievable fidelity $F_M$ using this strategy with the one obtained experimentally.

To mimic the quantum memory with this classical scheme, we provide Eve two possible strategies:
\begin{enumerate}
	\item[-] She can, in a first strategy, measure the totality of the input pulse and try to infer the qubit state. She then sends her guess in the transmitted as well as the stored pulse. To correctly reproduce the transmitted and output efficiencies, a loss $\delta$ can be adjusted in the output state. This strategy is depicted in the top part of figure {\ref{fideltrans}}.
	\item[-] In a second strategy, instead of measuring the whole input state she can simply transmit a part of the state with a beamsplitter of transmission $\eta$ to reproduce the transmitted pulse, and measure the reflected state to reproduce the output pulse. As the fidelity of the transmitted state is generally not expected to be perfect, noise must be added to the state by sending a completely mixed state with probability $q$.
\end{enumerate}

These two strategies are complementary in the sense that the first one could lead to a too high output fidelity, at the cost of the transmitted state fidelity, while the second strategy could have the opposite effect. Then, for each input pulse, Eve has the possibility to choose between these two strategies: the first one will be chosen with probability $p$, and the second one with probability $1-p$.

\subsubsection{Mathematical derivations}
\paragraph{Transmitted state:\\}
According to the previously depicted scheme, the fidelity and the efficiency of the transmitted state are:
\begin{eqnarray}
F_t&=&\frac{1}{\eta_t}\Big[p\eta_M^{(1)}F_M^{(1)}+(1-p)\eta\Big(\frac{1+q}{2}\Big)\Big]\\
\eta_t&=&p\eta_M^{(1)}+(1-p)\eta,
\end{eqnarray}
where $F_M^{(1)}=F_M(\mu,\eta_M^{(1)})$ and $F_M$ is given by equation (\ref{fidelbase}).
As $F_t$ and $\eta_t$ are fixed in our scheme, the two previous equations can be inverted to find expressions for $p$ and $\eta$:
\begin{eqnarray}
p&=&\frac{\eta_t}{\eta_M^{(1)}}\frac{\frac{1+q}{2}-F_t}{\frac{1+q}{2}-F_M^{(1)}}\label{condP}\\
\eta&=&\frac{\eta_t-p\eta_M^{(1)}}{1-p}\label{condEta}.
\end{eqnarray}

\paragraph{Output state:\\}
Similarly, for the output state:
\begin{eqnarray}
F_M&=&\frac{1}{\eta_{M}}\Big[p\delta\eta_M^{(1)}F_M^{(1)}+(1-p)(1-\eta)\eta_M^{(2)}F_M^{(2)}\Big]\\
\eta_{M}&=&p\delta\eta_M^{(1)}+(1-p)(1-\eta)\eta_M^{(2)},
\end{eqnarray}
with $F_M^{(2)}=F_M((1-\eta)\mu,\eta_M^{(2)})$. Given that $\eta_M$ is fixed we have
\begin{equation}
\eta_M^{(2)}=\frac{\eta_{M}-p\delta\eta_M^{(1)}}{(1-p)(1-\eta)}\label{condEta2}.
\end{equation}
The strategy of Eve is to search for the maximal fidelity $F_M$ by optimizing over the parameters $\eta_M^{(1)}$, $\delta$ and $q$.

\paragraph{Existence of a solution:\\}
We first show that a solution reproducing the measured efficiencies and the transmitted fidelity always exists in our model, which might not maximize the output state fidelity however. If we set $p=0$, $q=2F_t-1$, $\eta=\eta_t$ and $\eta_M^{(2)}=\eta_M$, the setup can fulfill all the conditions (\ref{condP}), (\ref{condEta}) and (\ref{condEta2}), so that there is always a possibility to reproduce the conditional fidelity of the transmitted state $F_t$ and the efficiencies $\eta_t$ and $\eta_M$ of the transmitted and output states respectively. It has to be noticed that in this case, only the second strategy is used ($p=0$).

\begin{figure}[!h]
\begin{center}
\includegraphics[width=12cm]{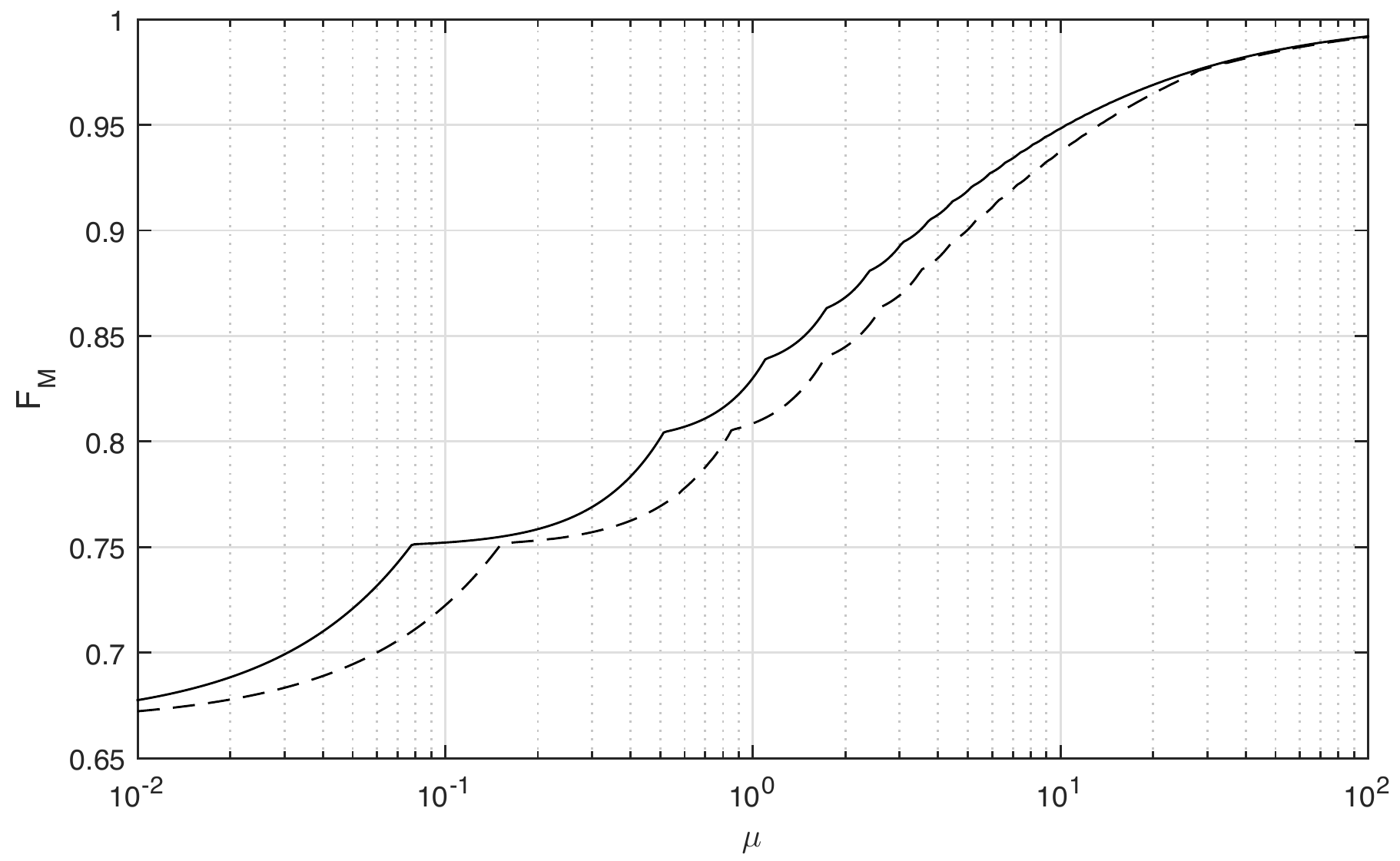}
\caption{Maximum achievable conditional fidelity of the output state $F_M$ by using the two measure-and-prepare strategies discussed here. Solid line: use of strategy presented in \cite{Gundogan2012} (see eq. \ref{fidelbase}) with $\eta=3.85\%$. Dashed line: use of the strategy described here, for $F_t=97.2\%$, $\eta_t = 29.6\%$ and $\eta_M=3.85\%$.}
\label{crit_newgen}
\end{center}
\end{figure}

\paragraph{Maximum "classical" fidelity based on our memory parameters:\\}
In figure \ref{crit_newgen}, we show the maximized conditional fidelity as a function of the mean photon number $\mu$ of the coherent state, for $F_t=97.2\%$, $\eta_t = 29.6\%$ and $\eta_M=3.85\%$. The figure reveals that the maximum achievable fidelity with the classical strategy is lower by taking into account the transmitted state.

\newpage
\section*{References}
\providecommand{\newblock}{}

\bibliographystyle{iopart-num}

\begin{thebibliography}{10}
\expandafter\ifx\csname url\endcsname\relax
  \def\url#1{{\tt #1}}\fi
\expandafter\ifx\csname urlprefix\endcsname\relax\def\urlprefix{URL }\fi
\providecommand{\eprint}[2][]{\url{#2}}

\bibitem{Gisin2007a}
Gisin N and Thew R 2007 {\em Nat Photon\/} {\bf 1} 165--171
  \urlprefix\url{http://dx.doi.org/10.1038/nphoton.2007.22}

\bibitem{Kimble2008}
Kimble H~J 2008 {\em Nature\/} {\bf 453} 1023--1030
  \urlprefix\url{http://dx.doi.org/10.1038/nature07127}

\bibitem{Briegel1998}
Briegel H~J, D\"ur W, Cirac J~I and Zoller P 1998 {\em Phys. Rev. Lett.\/} {\bf
  81} 5932--5935

\bibitem{Duan2001}
Duan L~M, Lukin M~D, Cirac J~I and Zoller P 2001 {\em Nature\/} {\bf 414}
  413--418 \urlprefix\url{http://dx.doi.org/10.1038/35106500}

\bibitem{Sangouard2011}
Sangouard N, Simon C, de~Riedmatten H and Gisin N 2011 {\em Rev. Mod. Phys.\/}
  {\bf 83} 33--80
  \urlprefix\url{http://link.aps.org/doi/10.1103/RevModPhys.83.33}

\bibitem{Lvovsky2009}
Lvovsky A~I, Sanders B~C and Tittel W 2009 {\em Nat Photon\/} {\bf 3} 706--714
  \urlprefix\url{http://dx.doi.org/10.1038/nphoton.2009.231}

\bibitem{Simon2010}
Simon C, Afzelius M, Appel J, Boyer de~la Giroday A, Dewhurst S~J, Gisin N, Hu
  C~Y, Jelezko F, Kr\"oll S, M\"uller J~H, Nunn J, Polzik E~S, Rarity J~G,
  De~Riedmatten H, Rosenfeld W, Shields A~J, Sk\"old N, Stevenson R~M, Thew R,
  Walmsley I~A, Weber M~C, Weinfurter H, Wrachtrup J and Young R~J 2010 {\em
  The European Physical Journal D - Atomic, Molecular, Optical and Plasma
  Physics\/} {\bf 58} 1--22
  \urlprefix\url{http://dx.doi.org/10.1140/epjd/e2010-00103-y}

\bibitem{Bussieres2013}
Bussi\`{e}res F, Sangouard N, Afzelius M, de~Riedmatten H, Simon C and Tittel W
  2013 {\em Journal of Modern Optics\/} {\bf 60} 1519--1537
  \urlprefix\url{http://dx.doi.org/10.1080/09500340.2013.856482}

\bibitem{Collins2007}
Collins O~A, Jenkins S~D, Kuzmich A and Kennedy T~A~B 2007 {\em Phys. Rev.
  Lett.\/} {\bf 98} 060502--4
  \urlprefix\url{http://link.aps.org/abstract/PRL/v98/e060502}

\bibitem{Simon2007}
Simon C, de~Riedmatten H, Afzelius M, Sangouard N, Zbinden H and Gisin N 2007
  {\em Phys. Rev. Lett.\/} {\bf 98} 190503
  \urlprefix\url{http://link.aps.org/abstract/PRL/v98/e190503}

\bibitem{Nunn2008}
Nunn J, Reim K, Lee K~C, Lorenz V~O, Sussman B~J, Walmsley I~A and Jaksch D
  2008 {\em Phys. Rev. Lett.\/} {\bf 101} 260502--4
  \urlprefix\url{http://link.aps.org/abstract/PRL/v101/e260502}

\bibitem{Usmani2010}
Usmani I, Afzelius M, de~Riedmatten H and Gisin N 2010 {\em Nat Commun\/} {\bf
  1} 12-- \urlprefix\url{http://dx.doi.org/10.1038/ncomms1010}

\bibitem{Bonarota2011a}
Bonarota M, Gou\"{e}t J~L~L and Chaneli\`{e}re T 2011 {\em New Journal of
  Physics\/} {\bf 13} 013013
  \urlprefix\url{http://stacks.iop.org/1367-2630/13/i=1/a=013013}

\bibitem{Higginbottom2012}
Higginbottom D~B, Sparkes B~M, Rancic M, Pinel O, Hosseini M, Lam P~K and
  Buchler B~C 2012 {\em Phys. Rev. A\/} {\bf 86} 023801

\bibitem{Zhou2015}
Zhou Z~Q, Hua Y~L, Liu X, Chen G, Xu J~S, Han Y~J, Li C~F and Guo G~C 2015 {\em
  Phys. Rev. Lett.\/} {\bf 115}(7) 070502
  \urlprefix\url{http://link.aps.org/doi/10.1103/PhysRevLett.115.070502}

\bibitem{Nicolas2014}
Nicolas A, Veissier L, Giner L, Giacobino E, Maxein D and Laurat J 2014 {\em
  Nat. Phot.\/} {\bf 8} 234--238

\bibitem{Sinclair2014}
Sinclair N, Saglamyurek E, Mallahzadeh H, Slater J~H, George M, Ricken R,
  Hedges M~P, Oblak D, Simon C, Sohler W and Tittel W 2014 {\em Phys. Rev.
  Lett.\/} {\bf 113} 053603
  \urlprefix\url{https://journals.aps.org/prl/accepted/fd07aYcaZ0d1114d423415468e37be0c2fae8946a}

\bibitem{Tittel2010b}
Tittel W, Afzelius M, Chaneli\`{e}re T, Cone R, Kr\"{o}ll S, Moiseev S and
  Sellars M 2010 {\em Laser \& Photonics Reviews\/} {\bf 4} 244--267
  \urlprefix\url{http://dx.doi.org/10.1002/lpor.200810056}

\bibitem{RiedmattenAfzeliusChapter2015}
Afzelius M and de~Riedmatten H 2015 Quantum light storage in solid state atomic
  ensembles {\em Engineering the Atom-Photon Interaction: Controlling
  Fundamental Processes with Photons, Atoms and Solids\/} Nano-Optics and
  Nanophotonics ed Predojevi\'{c} A and Mitchell M~W (Springer International
  Publishing) pp 241--268 ISBN 9783319192307
  \urlprefix\url{https://books.google.es/books?id=Fz5XrgEACAAJ}

\bibitem{Thiel2014}
Thiel C~W, Sinclair N, Tittel W and Cone R~L 2014 {\em Phys. Rev. Lett.\/} {\bf
  113} 160501

\bibitem{Longdell2005}
Longdell J~J, Fraval E, Sellars M~J and Manson N~B 2005 {\em Phys. Rev.
  Lett.\/} {\bf 95} 063601--4
  \urlprefix\url{http://link.aps.org/abstract/PRL/v95/e063601}

\bibitem{Heinze2013}
Heinze G, Hubrich C and Halfmann T 2013 {\em Phys. Rev. Lett.\/} {\bf 111}(3)
  033601 \urlprefix\url{http://link.aps.org/doi/10.1103/PhysRevLett.111.033601}

\bibitem{Zhong2015}
Zhong M, Hedges M~P, Ahlefeldt R~L, Bartholomew J~G, Beavan S~E, Wittig S~M,
  Longdell J~J and Sellars M~J 2015 {\em Nature\/} {\bf 517} 177--180 ISSN
  0028-0836 \urlprefix\url{http://dx.doi.org/10.1038/nature14025}

\bibitem{Hedges2010}
Hedges M~P, Longdell J~J, Li Y and Sellars M~J 2010 {\em Nature\/} {\bf 465}
  1052--1056 \urlprefix\url{http://dx.doi.org/10.1038/nature09081}

\bibitem{Sabooni2013}
Sabooni M, Li Q, Kr\"{o}ll S S and Rippe L 2013 {\em Phys. Rev. Lett.\/} {\bf
  110} 133604
  \urlprefix\url{http://link.aps.org/doi/10.1103/PhysRevLett.110.133604}

\bibitem{Jobez2014}
Jobez P, Usmani I, Timoney N, Laplane C, Gisin N and Afzelius M 2014 {\em New
  Journal of Physics\/} {\bf 16} 083005
  \urlprefix\url{http://stacks.iop.org/1367-2630/16/i=8/a=083005}

\bibitem{Gundogan2013}
G\"{u}ndo\u{g}an M, Mazzera M, Ledingham P~M, Cristiani M and de~Riedmatten H
  2013 {\em New Journal of Physics\/} {\bf 15} 045012
  \urlprefix\url{http://iopscience.iop.org/1367-2630/15/4/045012}

\bibitem{Timoney2013}
Timoney N, Usmani I, Jobez P, Afzelius M and Gisin N 2013 {\em Phys. Rev. A\/}
  {\bf 88} 022324
  \urlprefix\url{http://link.aps.org/doi/10.1103/PhysRevA.88.022324}

\bibitem{Gundogan2015}
G\"undo\ifmmode~\breve{g}\else \u{g}\fi{}an M, Ledingham P~M, Kutluer K,
  Mazzera M and de~Riedmatten H 2015 {\em Phys. Rev. Lett.\/} {\bf 114}(23)
  230501 \urlprefix\url{http://link.aps.org/doi/10.1103/PhysRevLett.114.230501}

\bibitem{Jobez2015}
Jobez P, Laplane C, Timoney N, Gisin N, Ferrier A, Goldner P and Afzelius M
  2015 {\em Phys. Rev. Lett.\/} {\bf 114}(23) 230502
  \urlprefix\url{http://link.aps.org/doi/10.1103/PhysRevLett.114.230502}

\bibitem{Clausen2012}
Clausen C, Bussi\`{e}res F, Afzelius M and Gisin N 2012 {\em Phys. Rev.
  Lett.\/} {\bf 108} 190503
  \urlprefix\url{http://link.aps.org/doi/10.1103/PhysRevLett.108.190503}

\bibitem{Gundogan2012}
G\"{u}ndo\u{g}an M, Ledingham P~M, Almasi A, Cristiani M and de~Riedmatten H
  2012 {\em Phys. Rev. Lett.\/} {\bf 108} 190504
  \urlprefix\url{http://link.aps.org/doi/10.1103/PhysRevLett.108.190504}

\bibitem{Zhou2012}
Zhou Z~Q, Lin W~B, Yang M, Li C~F and Guo G~C 2012 {\em Phys. Rev. Lett.\/}
  {\bf 108} 190505--
  \urlprefix\url{http://link.aps.org/doi/10.1103/PhysRevLett.108.190505}

\bibitem{Afzelius2009a}
Afzelius M, Simon C, de~Riedmatten H and Gisin N 2009 {\em Phys. Rev. A\/} {\bf
  79} 052329 \urlprefix\url{http://link.aps.org/abstract/PRA/v79/e052329}

\bibitem{Maudsley1986}
Maudsley A 1986 {\em Journal of Magnetic Resonance\/} {\bf 69} 488 -- 491 ISSN
  0022-2364
  \urlprefix\url{http://www.sciencedirect.com/science/article/pii/0022236486901605}

\bibitem{Souza2011}
Souza A~M, \'Alvarez G~A and Suter D 2011 {\em Phys. Rev. Lett.\/} {\bf
  106}(24) 240501
  \urlprefix\url{http://link.aps.org/doi/10.1103/PhysRevLett.106.240501}

\bibitem{AliAhmed2013a}
{Ali Ahmed} M~A, \'{A}lvarez G~A and Suter D 2013 {\em Phys. Rev. A\/} {\bf 87}
  042309

\bibitem{Silver1985}
Silver M~S, Joseph R~I and Hoult D~I 1985 {\em Phys. Rev. A\/} {\bf 31}(4)
  2753--2755 \urlprefix\url{http://link.aps.org/doi/10.1103/PhysRevA.31.2753}

\bibitem{Jobez2015b}
Jobez P, Timoney N, Laplane C, Etesse J, Gisin N, Ferrier A, Goldner P and
  Afzelius M 2015 {\em arXiv:1512.02936\/}
  \urlprefix\url{http://arxiv.org/abs/1512.02936}

\bibitem{Koenz2003}
K\"onz F, Sun Y, Thiel C~W, Cone R~L, Equall R~W, Hutcheson R~L and Macfarlane
  R~M 2003 {\em Phys. Rev. B\/} {\bf 68} 085109
  \urlprefix\url{http://link.aps.org/abstract/PRB/v68/e085109}

\bibitem{Ferrier2015}
Ferrier A, Tumino B and Goldner P 2015 {\em Journal of Luminescence\/}  -- ISSN
  0022-2313
  \urlprefix\url{http://www.sciencedirect.com/science/article/pii/S002223131500407X}

\bibitem{Yano1991}
Yano R, Mitsunaga M and Uesugi N 1991 {\em Opt. Lett.\/} {\bf 16} 1884--1886
  \urlprefix\url{http://ol.osa.org/abstract.cfm?URI=ol-16-23-1884}

\bibitem{Yano1992a}
Yano R, Mitsunaga M and Uesugi N 1992 {\em J. Opt. Soc. Am. B\/} {\bf 9}
  992--997 \urlprefix\url{http://josab.osa.org/abstract.cfm?URI=josab-9-6-992}

\bibitem{Massar1995}
Massar S and Popescu S 1995 {\em Phys. Rev. Lett.\/} {\bf 74}(8) 1259--1263
  \urlprefix\url{http://link.aps.org/doi/10.1103/PhysRevLett.74.1259}

\bibitem{Specht2011}
Specht H~P, Nolleke C, Reiserer A, Uphoff M, Figueroa E, Ritter S and Rempe G
  2011 {\em Nature\/} {\bf 473} 190--193
  \urlprefix\url{http://dx.doi.org/10.1038/nature09997}

\bibitem{Nielsen2000}
Nielsen M~A and Chuang I~L 2000 {\em Quantum Computation and Quantum
  Information\/} (Cambridge)

\end{thebibliography}

\end{document}